\documentclass[printer]{aa}%
\usepackage{graphicx}
\usepackage[round]{natbib}
\usepackage{fancyhdr}
\usepackage{amsmath}
\usepackage{subfig}

\newcommand{\on}{[O\,{\sc i}]}

\newcommand{\snn}{[S\,{\sc ii}]}
\newcommand{\cf}{C\,{\sc iv}}
\newcommand{\lx}{$L_{\rm X}$}
\newcommand{\luv}{$L_{\rm FUV}$}
\newcommand{\mic}{$\rm \mu m$}
\newcommand{\apo}{$a_{\rm pow}$}
\newcommand{\ami}{$a_{\rm min}$} 
\newcommand{\eps}{$\epsilon$}

\newcommand{\lac}{$L_{\rm acc}$}
\newcommand{\lsum}{$L_{\rm SUM}$}
\begin{document}
\title{\on\,disk emission in the Taurus star forming region.}
\author{G.~Aresu\inst{1,2}
  \and I.~Kamp\inst{1}
  \and R.~Meijerink\inst{1,3}
  \and M.~Spaans\inst{1}
  \and S.~Vicente\inst{1}
  \and L.~Podio\inst{5,6}
  \and P.~Woitke\inst{4}
  \and F.~Menard\inst{5,7}
  \and W.-F.~Thi\inst{5}
  \and M.~G\"{u}del\inst{8}
  \and A.~Liebhart\inst{8}}
\institute{Kapteyn Astronomical Institute, Postbus 800, 9700 AV Groningen, The Netherlands
 \and INAF, Osservatorio Astronomico di Cagliari, Via della Scienza 5, Selargius (CA), 09047, Italy
 \and Sterrewacht Leiden, Leiden University, P.O. Box 9513, 2300 RA Leiden, The Netherlands
 \and SUPA, School of Physics and Astronomy, University of St. Andrews KY16 9SS, UK
 \and UJF-Grenoble, CNRS-INSU, Institute de Plan\`etologie et d'Astrophysique (IPAG) UMR 5274
 \and INAF - Osservatorio Astrofisico di Arcetri, Largo E. Fermi 5, 50125, Florence, Italy
 \and UMI-FCA (UMI 3386), CNRS/INSU France and Universidad de Chile, 1058 Santiago, Chile  
 \and University of Vienna, Department of Astronomy, T\"urkenschanzstrasse 17, 1180 Vienna, Austria}
 \date{Received ? / Accepted ?}

\abstract{The structure of protoplanetary disks is thought to be linked to the temperature and chemistry of their dust and gas. Whether the disk is flat or flaring depends on the amount of radiation that it absorbs at a given radius, and on the efficiency with which this is converted into thermal energy. The understanding of these heating and cooling processes is crucial to provide a reliable disk structure for the interpretation of dust continuum emission and gas line fluxes. Especially in the upper layers of the disk, where gas and dust are thermally decoupled, the infrared line emission is strictly related to the gas heating/cooling processes.}{We aim to study the thermal properties of the disk in the oxygen line emission region, and to investigate the relative importance of X-ray (1-120 \AA) and far-UV  radiation (FUV, 912-2070 \AA) for the heating balance there.}{We use \on\,63 \mic\,line fluxes observed in a sample of protoplanetary disks of the Taurus/Auriga star forming region and compare it to the model predictions presented in our previous work. The data were obtained with the PACS instrument on board the Herschel Space Observatory as part of the Herschel Open Time Key Program GASPS (GAS in Protoplanetary diskS), published in Howard et al. (2013).}{Our theoretical grid of disk models can reproduce the \on\,absolute fluxes and predict a correlation between \on\,and the sum \lx+\luv. The data show no correlation between the \on\,line flux and the X-ray luminosity, the FUV luminosity or their sum.}{The data show that the FUV or X-ray radiation has no notable impact on the region where the \on\,line is formed. This is in contrast with what is predicted from our models. Possible explanations are that the disks in Taurus are less flaring than the hydrostatic models predict, and/or that other disk structure aspects that were left unchanged in our models are important. Disk models should include flat geometries, varying parameters such as outer radius, dust settling, and the dust-to-gas mass ratio, which might play an equally important role for the \on\,emission. To improve statistics and draw more robust conclusions on the thermal processes that dominate the atmosphere of protoplanetary disks surrounding T\,Tauri stars, more \luv\,and \lx\,measurements are needed. High spatial and spectra resolution data is required to disentangle the fraction of \on\,flux emitted by the disk in outflow sources.}

\maketitle

\section{Introduction}

Planet formation is strongly linked to the physical properties of the parent disk. Important constraints on the timescale for the gas accretion of giant planets are posed by photoevaporation models. The results of such models are essential in order to estimate the mass loss rates, and hence the survival time of gas in disks (\citealt{Ale06,Erc08,GDH09}). The stellar radiation, especially in the high energy regime (E $>$ 6 eV), is responsible for the thermo-chemical conditions in the disk atmosphere, as it provides most of the energy that causes the gas temperature to exceed the dust temperature there \citep{Kam04,Jon04,Gla04}. However, the thermal processes that heat and shape protoplanetary disks are poorly constrained and can only be indirectly measured through cooling lines. One of the dominant cooling lines that can be used to understand these processes is the 63 micron line of neutral oxygen \citep{Gor08,Mei08,Woi09,Are12}.

T\,Tauri stars emit radiation at high energies, due both to chromospheric activity and accretion of disk material onto the stellar surface. The FUV luminosity between 7 and 10 eV ($\Delta\lambda$=1240-1770 \AA), has been measured by \citet{Yan12} for a sample of accreting sources in Taurus: they found values between 10$^{30}$ and few times 10$^{32}$ erg/s. The emission is in excess when compared to the stellar emission in the same energy band for non-accreting young stars of the same spectral type. This suggests that accretion is responsible for this emission, in which case it is caused by shocks created by the magnetic field that channels disk material toward the stellar surface \citep{Cal98,Val00}. EUV ($\Delta\lambda$= 120-912 \AA, $\Delta$E=13.6-100 eV) radiation is believed to mainly affect the upper disk surface at small radii, as the high cross section for absorption only allows penetration of small columns of $N_{\rm H}\sim 10^{19}$ cm$^{-2}$. The XEST survey \citep{Gue07} has shown that young stars are also active X-ray emitters, mainly due to chromospheric activity, and can reach luminosities between 10$^{29}$ and 10$^{31}$ erg/s. The high energy depositions ($>$0.01 $L_{*}$) and heating efficiencies ($\sim$ 30\%) of X-rays cause the tenuous disk atmosphere to heat up to temperatures of the order of a few thousand Kelvin \citep{Gla07,Nom07,Gor08,Erc08,Are11}. 

Recent observations, carried out with the Herschel Space Observatory toward the Taurus forming region, offer the chance to test model predictions on the thermal structure of the region where the \on\,63.2 \mic\,line is emitted. This line is predicted to arise from the disk atmosphere in the radial region between a few 10 AU and 200 AU \citep{Woi09,Are12}. The emission region is directly exposed to the stellar radiation and models suggest that FUV and X-ray radiation are the main heating agents there. PAH and dust photoelectric heating as well as Coulomb heating, cause the gas temperature to be of the order of $\sim$ 200-300 K \citep{Gor08,Mei08,Mei12}.

In this work, we explore possible correlations of the \on\,emission with X-ray luminosity and FUV luminosity, and compare the \citet{Are12} model predictions for the \on\,63 \mic\,emission with data collected within the GASPS (GAS in Protoplanetary DiskS, P.I. Dent) Open Time Key Program, taken with the PACS instrument on board the Herschel Space Observatory \citep{Den13}.

In the following, we make the hypothesis that most of the \on\,emission is produced in the disk. Outflow sources, that have on average higher accretion rates, will then produce more FUV radiation and thus stronger FUV illumination of the disk surface and stronger line emission. 
 
In Sect.\ref{obser} we present the collected observational data set and in Sect.\ref{models} we explain the main findings of the models studied in \citet{Are12}. In Sect.\ref{results} we show the results of the comparison between model predictions and observations, these will be discussed in Sect.\ref{disc}. Conclusions and remarks about future work are summarised in Sect.\ref{fut}.

\section{Observations}
\label{obser}
In Table \ref{sample} we list the sources studied in this paper, together with the observed \on\,fluxes, and the collected X-ray and FUV luminosities with references. The Taurus star-forming region contains a rich population of pre-main sequence stars, with an age between 1-3 Myr. The sources in this sample have spectral types G, K or M, and the majority of these are Class II objects. Below we describe the origin of each observed quantity listed in Table \ref{count}.

\begin{table*}[ht!]
 \centering
 \begin{tabular}{l|c|c|c||c|c|c}
 \hline
 \hline
 Name & Class & \on\,63 $\mu$m &  \on\,63 $\mu$m  & $L_{\rm FUV}$ & $L_{\rm X}$ & Outflow\\
   &  &  [1e-17 W/m$^2$] & disk only& [1e30 erg/s] &              [1e30 erg/s] & [YES/NO] \\
 \hline
AATau  &II&   2.2$\pm$0.2& 0.60 & 28.82$^1$&   1.24$_{\, 1.11}^{\, 1.36}$ & Y \\
 \hline
BPTau  &II&   0.10$\pm$0.03 & & 58.64$^1$&   1.36$_{\, 1.35}^{\, 1.40}$ & N \\
 \hline
CITau  &II&   3.3$\pm$0.5& & 13.32$^1$&   0.19$_{\, 0.16}^{\, 0.89}$ & N \\
 \hline
CWTau  &II&   7.2$\pm$0.4 & 0.82 & 111.13$^2$&   2.84$_{\, 0.28}^{\, 4.00}$ & Y \\
 \hline
CXTau  &T&   0.7$\pm$0.3 & & 0.68$^1$&   - & N \\
 \hline
CYTau  &II&   1.2$\pm$0.4 & &13.32$^1$&   0.13$_{\, 0.13}^{\, 0.29}$ & N \\
 \hline
DETau  &II&   0.7$\pm$0.6 & &30.41$^1$&   - & N \\
 \hline
DFTau  &II&   6.1$\pm$0.6 &  &9.95$^1$&   - & Y \\
 \hline
DGTau  &II& 134.00$\pm$17.0 & 4.70 & 318.04$^2$&   0.55$_{\, 0.39}^{\, 0.78}$ & Y \\
 \hline
DHTau  &II&   $<$1.35 &   & -  &8.46$_{\, 8.23}^{\, 8.64}$ & N \\
 \hline
DKTau  &II&   1.6$\pm$0.3  &  &18.58$^2$&   0.92$_{\, 0.87}^{\, 0.96}$ & N \\
 \hline
DLTau  &II&   2.2$\pm$0.2  &  &13.97$^1$&   - & Y \\
 \hline
DMTau  &T&   0.7$\pm$0.2 & & 58.28$^1$&   2.00$_{\, 1.41}^{\,2.83}$ & N \\
 \hline
DNTau  &T&   0.6$\pm$0.2 &  & 6.52$^1$&   1.15$_{\, 1.14}^{\, 1.17}$ & N \\
 \hline
DOTau  &II&   7.1$\pm$1.0  & 1.45 &470.60$^1$&   0.24$_{\, 0.17}^{\, 0.34}$ & Y \\
 \hline
DPTau  &II&  14.8$\pm$1.3  & 0.57 &96.29$^1$&   0.10$_{\, 0.04}^{\, 0.18}$ & Y \\
 \hline
DQTau  &II&   2.1$\pm$0.4  & &  0.82$^2$&   - & N \\
 \hline
DSTau  &II&   0.9$\pm$0.2  & & 49.96$^1$&   - & N \\
 \hline
FFTau  &III&   $<$1.01 &   & -&  0.80$_{\, 0.69}^{\,1.12}$ & N \\
 \hline
FMTau  &II&   1.0$\pm$0.2 &  & 4.89$^1$&   0.53$_{\, 0.51}^{\, 0.56}$ & N \\
 \hline
FOTau  &T&   1.20$\pm$0.5  &  & -&   0.06$_{\, 0.05}^{\,0.52}$ & N \\
 \hline
FQTau  &II&   $<$0.92 &   & - & 0.12$_{\, 0.05}^{\,0.83}$ & N \\
 \hline
FSTau-A&II-FS&  35.8$\pm$0.5 & & -&   3.22$_{\, 3.09}^{\,3.36}$ & Y \\
 \hline
FXTau  &II&   $<$1.38 &   & - & 0.50$_{\, 0.39}^{\, 2.36}$ & N \\
 \hline
GGTau  &II&   5.1$\pm$0.4  & & 10.51$^2$&   - & N \\
 \hline
GHTau  &II&   $<$0.85&   & - & 0.11$_{\, 0.10}^{\,0.12}$ & N \\
 \hline
GI-KTau&II &   3.1$\pm$1.4 & &  10.86$^1$&   0.83$_{\, 0.73}^{\, 1.06}$ & N \\
 \hline
GMAur  &T&   2.4$\pm$0.5  & & 28.24$^1$&   1.60$_{\, 1.13}^{\, 2.26}$ & N \\
 \hline
GOTau  &II&   $<$5.38&   & - & 0.25$_{\, 0.22}^{\,0.36}$ & N \\
 \hline
HBC358 &III&   $<$1.4 &   &- &  0.38$_{\, 0.37}^{\, 0.44}$ & N \\
 \hline
HKTau  &II&   3.4$\pm$0.2   &  & -&   0.08$_{\, 0.06}^{\,0.12}$ & N \\
 \hline
HLTau  &I &  51.3$\pm$0.5  & &   -&   3.84$_{\, 3.22}^{\, 4.73}$ & Y \\
 \hline
HNTau  &II&   4.1$\pm$0.2  & 0.56 &  21.29$^1$&   0.32$_{\, 0.23}^{\, 0.45}$ & Y \\
 \hline
HOTau  &II&   $<$1.03 &  & -&   0.05$_{\, 0.04}^{\, 0.05}$ & N \\
 \hline
Haro6-13&II&   7.0$\pm$0.5 & &  -&   0.80$_{\, 0.14}^{\, 0.91}$ & Y \\
 \hline
IPTau  &T&   0.6$\pm$0.2  &  & 4.05$^1$&   - & N \\
 \hline
IQTau  &II&   1.5$\pm$0.3  &  & -&   0.42$_{\,0.33}^{\, 1.17}$ & N \\
 \hline
IRAS043&II &   4.9$\pm$0.2  &  & -&   0.40$_{\, 0.37}^{\, 0.50}$ & Y \\
 \hline
LkCa15 &T&   1.0$\pm$0.2   & &  4.45$^3$&   - & N \\
 \hline
RWAur  &II&  15.4$\pm$0.5  &  & -&   1.60$_{\, 1.13}^{\,2.26}$ & Y \\
 \hline
RYTau  &T?&  10.5$\pm$0.5  & 3.80 & 1042.56$^1$&   5.52$_{\, 4.82}^{\, 6.38}$ & Y \\
\hline
SUAur  &II&   8.6$\pm$0.3  & 2.51 & 127.42$^1$&   9.46$_{\, 8.42}^{\, 9.70}$ & Y \\
 \hline
UYAur  &II&  31.4$\pm$0.4  & 2.16 & 27.73$^2$ &   0.40$_{\, 0.28}^{\, 0.57}$ & Y \\
 \hline
UZTau  &II&   4.5$\pm$1.4  &  & -&   0.89$_{\, 0.51}^{\, 1.35}$ & Y \\
 \hline
V710Tau&II&   1.0$\pm$0.6  &  & -&   1.38$_{\, 1.32}^{\, 1.49}$ & N \\
 \hline
V773Tau&II&   6.5$\pm$0.3  &  & -&   9.49$_{\, 9.39}^{\, 9.54}$ & Y \\
 \hline
V819Tau&II&   $<$0.898   &  & -&   2.44$_{\, 2.33}^{\, 2.61}$ & N \\
 \hline
XZTau  &II &  36.1$\pm$0.09  &  & -&   0.96$_{\, 0.86}^{\,1.12}$ & Y \\
 \hline
 \end{tabular}
\caption{\small Sources analysed in this work and their properties: Class, \on\,flux, \lx\,and \luv, and presence of an optical jet/outflow. The oxygen fluxes listed in the fourth column are the estimated disk contribution to the total \on\,flux as explained in 5.2.   The class of the objects are taken from \citet{And05}, X-ray luminosities between 0.3 and 10 kev are taken from \citet{Gue07,Gue10}, FUV luminosities with superscript 1 are retrieved from \citet{Yan12} (errors are estimated to be $\pm$30\%), while the ones with superscript 2 and 3 are calculated using the accretion luminosity retrieved from \citet{Gul98}, and \citet{Ing09} respectively (in this case the error is the one associated with the correlation: 0.38 dex). These values are then scaled by a factor 4.25 to estimate the luminosity in the 6-13.6 eV range. Outflow sources are objects for which extended emission in the optical, associated to a jet, has been observed. The \on\,emission from these objects has not been included in the primary analysis performed in this work.}
 \label{sample}
\end{table*}

\begin{table*}[t!]
\centering
\begin{tabular}{c|c|c||c|c|c|c}
\hline
\hline
Sample & \on\,detections & Upper limits & No outflow & \lx & \luv & \lx\,and \luv \\
\hline
 48 & 39 & 9 & 29 & 22 & 17 & 9\\
\hline
\end{tabular}
\caption{\small Taurus sources from GASPS. \lx\,is measured between 0.3 and 10 keV, \luv\,is measured between 7 and 10 eV. The last three columns refer only to sources for which no outflow emission was detected.}
\label{count}
\end{table*}
\subsection{Oxygen line fluxes}
The data reduction is described in \citet{How13}. The \on\,63 \mic\,line was detected in 39 out of 48 class II objects observed in Taurus, and upper limits could be measured for 9 more sources. Following Howard et al. (2013) we define jet/outflow sources those objects that have a jet imaged in H$\alpha$, \on\,$\lambda$6300 \AA, \snn\,$\lambda$6371 \AA\,or are associated with Herbig-Haro objects, and which show a high velocity molecular outflow, or a broad ($>$ 50 km s$^{−1}$), typically blue-shifted, emission line profile in \on\,$\lambda$6300  \AA (see e.g. \citealt{Har95}). These objects are labelled with a Y in Table \ref{sample}. \citet{Pod12} showed that for four of these sources (T Tau, DG Tau A, FS Tau and RW Aur) the oxygen emission at 63 \mic\,is spatially extended. They compared shock and disk model predictions for the fluxes of the \on\,63 \mic\,line, and found that these are likely dominated by jet/outflow emission. Following these arguments, in order to compare the data to our disk models, we only analyze \on\,fluxes from those sources in which no outflow emission has been detected (29 sources). From this sample we could retrieve the X-ray luminosity for 22 sources and the FUV luminosity for 17 sources. Both \lx\,and \luv\,were retrieved for 9 sources (Table \ref{count}).

\subsection{X-ray luminosities}
We collected the X-ray luminosity for 22 sources from the observations carried out with the XMM-Newton spacecraft toward the Taurus forming region, performed in the context of the XEST survey (P.I. M. Guedel). The X-ray luminosities range between 10$^{29}$ and 10$^{31}$ erg/s, these values and the associated errors are taken from \citet{Gue07} for all the sources, except DM\,Tau, GM\,Aur and HN\,Tau. For these objects, \lx\,was taken from \citet{Gue10}. In the latter case, as suggested by the authors, an error of $\pm\sqrt{2}$\lx\,is associated with the X-ray luminosity value to account for intrinsic variability, which is the dominant source of error. 
\subsection{FUV luminosities}
We could retrieve FUV luminosities for 13 sources from \citet{Yan12} (Table \ref{sample}), the associated errors for these values are of the order of $\sim$30\%. The observations were performed with the ACS camera and STIS spectrograph on board the Hubble Space Telescope. The FUV luminosity is obtained integrating in the 1240-1770 \AA\,range (7-10 eV) over the dominant line emission of, e.g., \cf 1459 \AA, Si\,{\sc iv} 1394 \AA, after continuum subtraction and correction for interstellar extinction using the law by \citealp{Car89} ($R_V=3.1$). The uncertainty in the A$_V$ and extinction law can contribute significantly to the error in the observed line fluxes. The authors assume an error in A$_V$ of $\sim$ 0.5 mag when no errors are available.
The FUV luminosity is correlated with the  accretion luminosity $L_{\rm acc}$. To extend the number of \luv\,measurements, we attempted to derive FUV luminosities for those objects in our sample that are not listed in \citet{Yan12}, using the correlations they provide for \lac\,and \luv (in units of solar luminosity):
\begin{equation}
\label{trial}
\rm{log}(L_{\rm FUV})  = -1.67 + 0.84\,\rm{log}(L_{\rm acc}) \\
\end{equation} 

\begin{figure}
 \centering
 \includegraphics[scale=0.43,angle=-90]{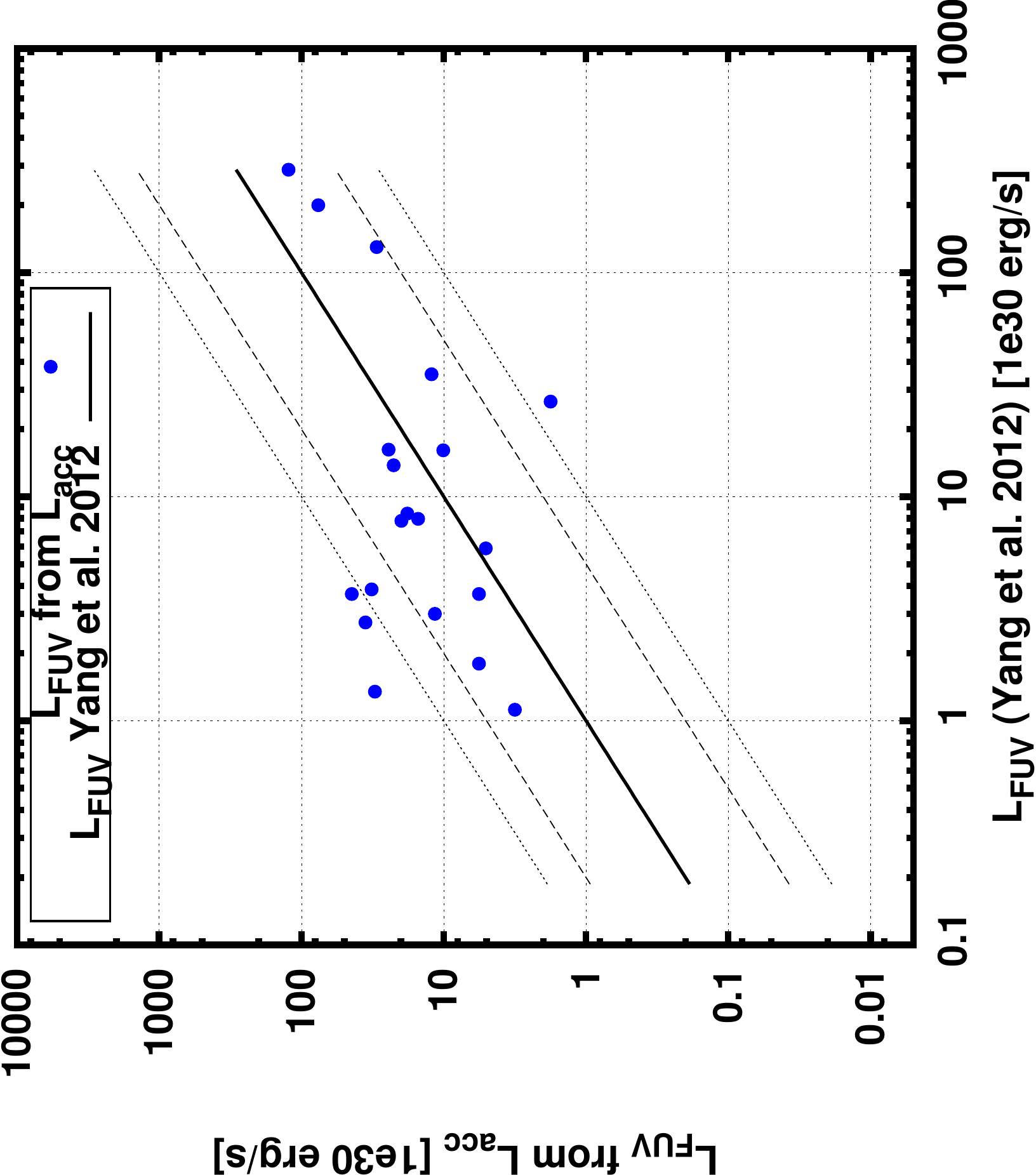}
 \caption{\small Comparison of the \luv\,luminosities obtained from \lac\, with the FUV luminosity listed in \citet{Yan12}. The solid line indicates a one-to-one ratio; dashed and dotted line encloses the region where the new \luv\,are a factor 5 and 10 higher/lower than \luv\,obtained by \citet{Yan12} with HST, respectively.} 
 \label{methods}
\end{figure}

The FUV luminosity obtained with this prescription accounts for chromospheric and accretion related emission.
To test this method we used \lac\,taken from \citet{Gul98} and \citet{Ing09} and compared the FUV luminosity found using Eq. (\ref{trial}), to the one provided by \citet{Yan12} for the sources in common.

\citet{Gul98} measure the accretion luminosity as follows: the excess flux in the energy range 2.4-3.9 eV (3200-5100 \AA) is estimated computing the relative veiling in the 2.8-3 eV (4100-4400 \AA) and 2.6-3.9 eV (3200-4800 \AA) bands, where clearly veiled absorption lines are available. The spectra are then corrected for the extinction. The accretion luminosity outside the 2.4-3.9 eV band is estimated considering a slab of constant temperature and density to model the accretion spots on the stellar surface. The statistical equilibrium is solved for hydrogen, and an escape probability method is applied to estimate the emitted flux. \citet{Gul98} find that the total excess flux, which is converted to accretion luminosity once the distance is known, is $\sim$3.5 times higher than the flux excess in the 2.4-3.9 eV band. They also note that the accretion luminosity is proportional to the luminosity in the dereddened U-band, and provide fit parameters for this relation. This is used by \citet{Ing09} to compute accretion luminosities for several other sources.
\begin{figure}[t]
\includegraphics[scale=0.41,angle=-90]{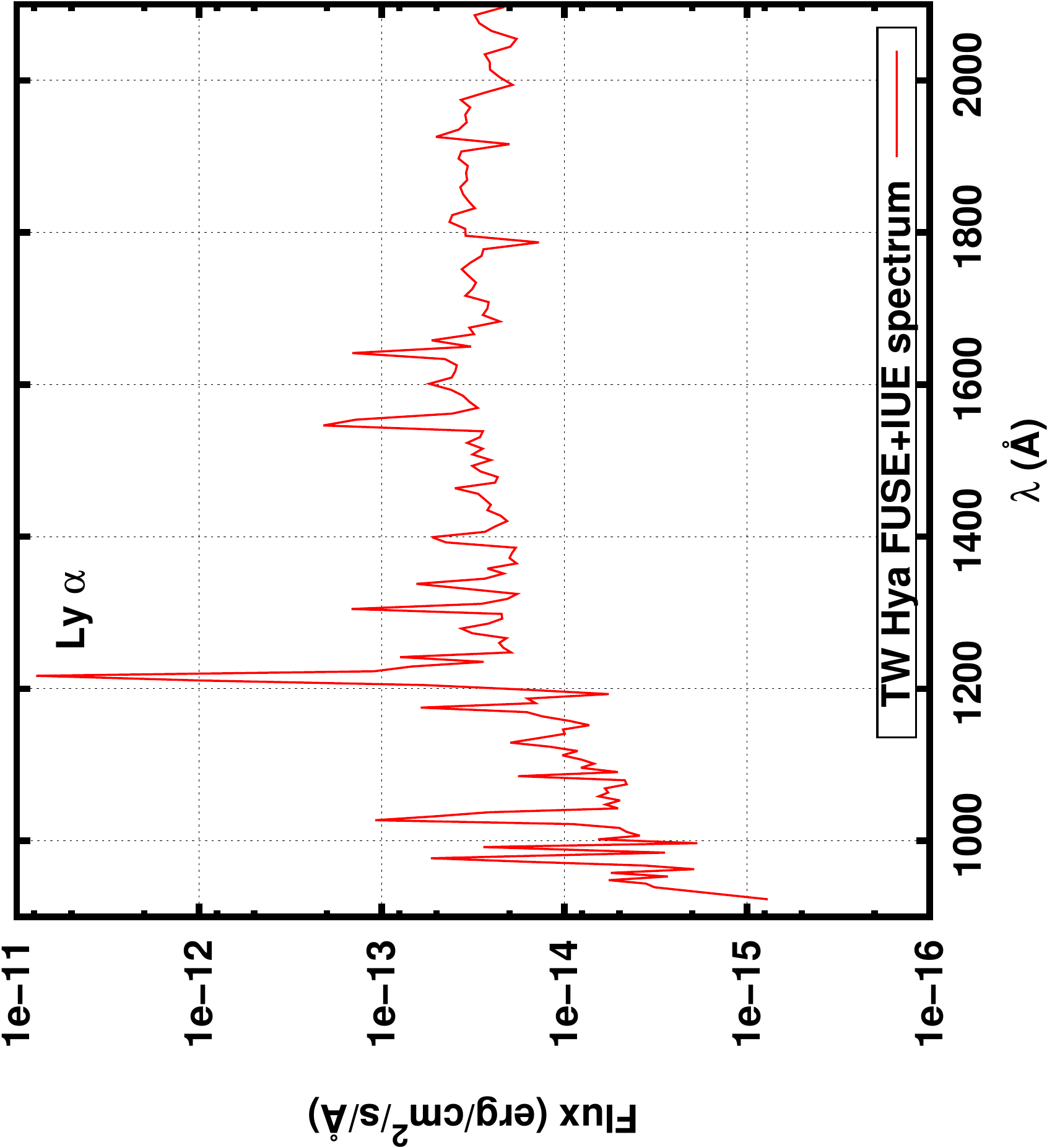}
\caption{TW Hya spectrum obtained co-adding IUE and FUSE data. This spectra is used to estimate the conversion factor between the flux in the 7-10 eV band and 6-13.6 eV band.}
\label{ly}
\end{figure}

The FUV emission in young stars is related to accretion, which is expected to be variable in time ($\sim$0.5 dex in days/months, \citealp{Ngu09}). Calculating the accretion luminosity considering a collection of photometry and spectral points, or from the correlation with the U broad band emission, likely guarantees a good estimate of the overall flux in the FUV band, causing variability to average out.

We were able to obtain \luv\,for 4 sources for which \on\,has been detected, extending the sample from 13 to 17. The error associated to the derived \luv\,is dominated by the mean scatter in the correlation with \lac\,(0.38 dex).
\begin{figure*}[t!]
 \centering
 \includegraphics[scale=0.4,angle=-90]{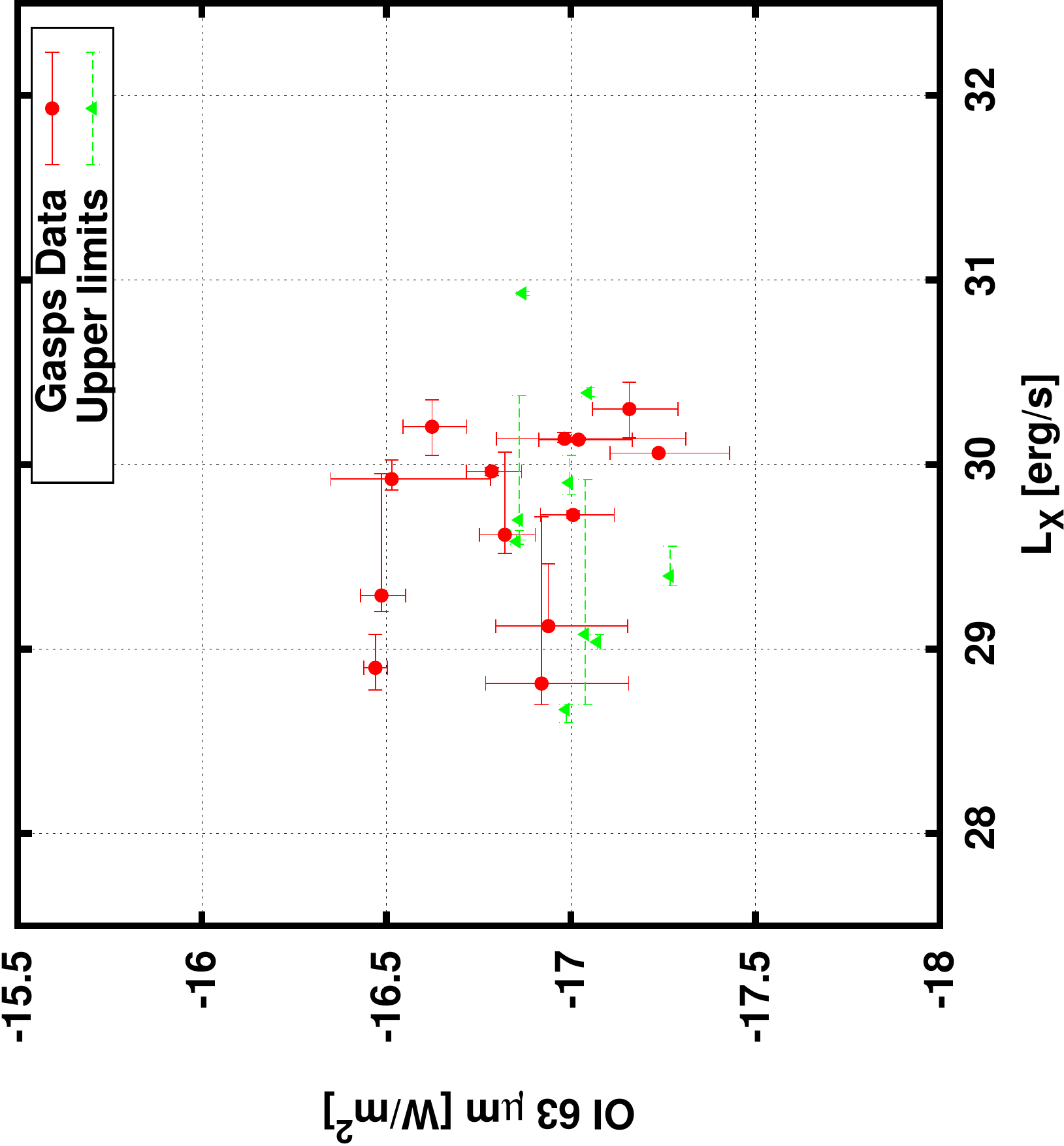} %
 \includegraphics[scale=0.4,angle=-90]{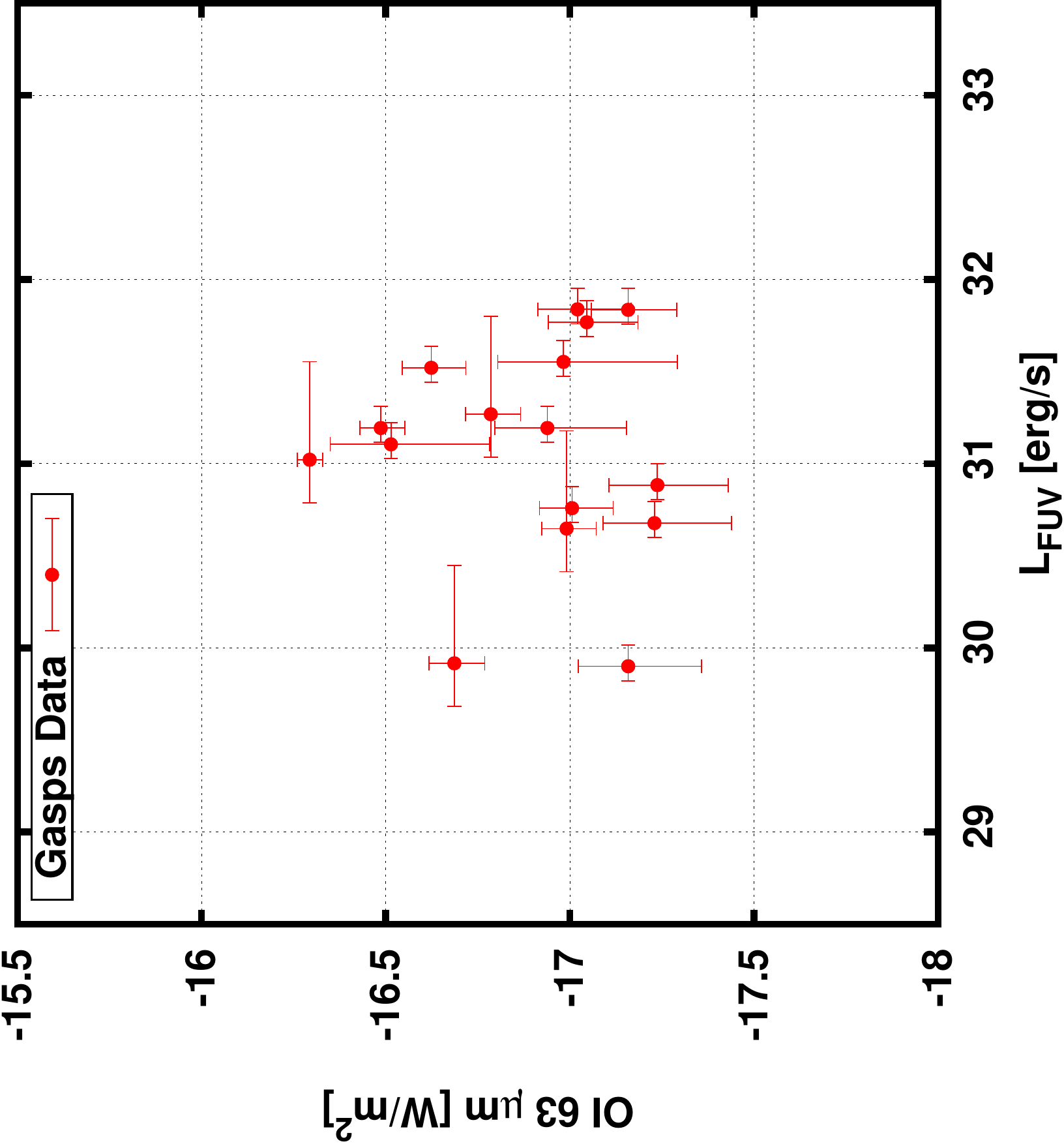}
 \caption{\small Red dots identify \on\,fluxes emitted from non-outflow sources, green diamonds are upper limits. Left-hand panel: flux of the \on\,63 \mic\,emission versus the X-ray luminosity. Right-hand panel: flux of the \on\,63 \mic\,emission versus the FUV luminosity.} 
 \label{oilxdata}
\end{figure*}
It is important to note that our models define the FUV luminosity in the range between 6 and 13.6 eV (92-250 nm), while \citet{Yan12} provides integrated fluxes from 7 to $\sim$10 eV (125-170 nm). We used a TW Hya spectrum composed of a collection of FUSE\footnote{http://archive.stsci.edu/fuse/ (6 data files)} (900-1190 \AA) and IUE\footnote{http://sdc.cab.inta-csic.es/cgi-ines/ (16 data files)} (1150-1980 \AA) data to calculate the luminosity ratio between the 6-13.6 eV and the 7-10 eV band. The spectrum, shown in Fig. \ref{ly}, was obtained first defining resolution dependent wavelength bins and then co-adding each dataset using the inverse square of the bin uncertainty as summation weight. We found a conversion factor of 4.25 for TW Hya:

\begin{equation}
L_{\rm FUV}^{(6-13.6\rm eV)} = 4.25\,L_{\rm FUV}^{(7-10\rm eV)}.
\end{equation}
We applied this conversion factor to all the other objects and from now on we will refer to \luv\, as the FUV luminosity between 6 and 13.6 eV. A very important contribution in the FUV band is given by the Ly$\alpha$ line emission which can carry up to 70-90$\%$ of the total FUV flux \citep{Sch12}. However, due to resonant scattering of neutral hydrogen and deuterium in the ISM, the calculation of the Ly$\alpha$ fluxes must rely on line profile reconstruction \citep{Fra12}, which is beyond the scope of this work. Moreover, \citet{Bet11} showed that Ly$\alpha$ is efficiently scattered through the atomic layers of protoplanetary disks by neutral hydrogen and dominates the energy budget over FUV continuum dee per in the disk, where the chemical environment is rich of molecules. In our models we find that the \on\,line is produced slightly above the H/H$_2$ transition but also that the emission is insensitive to the chemical conditions, but only sensitive to the temperature there, which is set by the interaction of the FUV continuum with PAHs and neutral carbon.

\section{Models}
\label{models}
In this work we use the results obtained in \citet{Are12}, where we used the thermo-chemical code ProDiMo \citep{Woi09,Are11} to calculate the \on\,63 \mic\,line fluxes for a grid of 240 models. The varying parameters in the grid are \lx\,(0, 10$^{29}$, 10$^{30}$,10$^{31}$, and 10$^{32}$ erg/s), \luv\,(10$^{29}$, 10$^{30}$,10$^{31}$, and 10$^{32}$ erg/s), minimum dust grain size \ami\,(0.1, 0.3, and 1 \mic), dust size distribution power law index \apo\,(2.5,3.5), and surface density distribution power law index \eps\,(1.0,1.5). The PAH abundance is 1\% with respect to the ISM abundance ($\epsilon_{\rm ISM}=3\times 10^{-7}$) and the dust to gas ratio is kept fixed at 0.01 throughout the whole disk. Following \citet{Woi09} we considered a turbulent Doppler value of 0.15 km/s \citep{Gui98,Sim00}.

As described in \citet{Are12}, among the parameters described above, the main effect on \on\,is caused by \lx\,and \luv. We then calculate the mean \on\,flux over each series of models with a given value of \lx\,and \luv. One series is composed of 12 models, which differ for values of \ami, \apo\, and \eps. The error bars accompanying the mean flux take into account a deviation of 2$\sigma$ in that sub-series of 12 models. We find that the behaviour of the line flux predicted from the models (also shown in Fig. 4) is not affected by the disk inclination.

We found that the \on\,line is optically thick ($\tau_{\rm line}>$10$^4$), hence sensitive to the gas temperature in the disk regions between r $\sim $10 and 100 AU with relative height z/r increasing from 0.2 to 0.6 (see Fig. 2 in Aresu et al., 2012). The main FUV related heating processes are photoelectric heating on PAHs, dust grains and carbon ionization heating. In all cases these processes release a few eV into the gas phase which are converted into kinetic energy of the gas. X-ray heating proceeds via Coulomb heating, which releases larger amounts of energy due to the fast electrons released in the X-ray ionization process. We found that depending on \luv\,and \lx\,luminosities the temperature in the \on\,heating region, and consequently the line flux, is controlled by FUV or X-ray radiation or both. 

In our grid of models we consider a single star, of spectral type G (see \citealt{Mei12}, Table 1). The spread in spectral types in Taurus is restricted to objects of spectral type G, K and M, for which the effective temperatures and bolometric luminosities agree reasonably well respect to our Sun-like model. This might influence the SED properties of such systems, but it does not play a role in the gas physics and chemistry in the upper layers, as this is regulated by high energy radiation. 

\section{Results}
\label{results}
We describe the results of the observations of oxygen emission in Taurus, and investigate the correlation between the oxygen fine-structure line at 63 \mic\,and \lx, \luv\,and their sum. We then compare the data results to the model predictions described in \citet{Are12}, to study the thermal properties of the region where \on\, is emitted.
\begin{figure*}
\centering
 \includegraphics[scale=0.4,angle=-90]{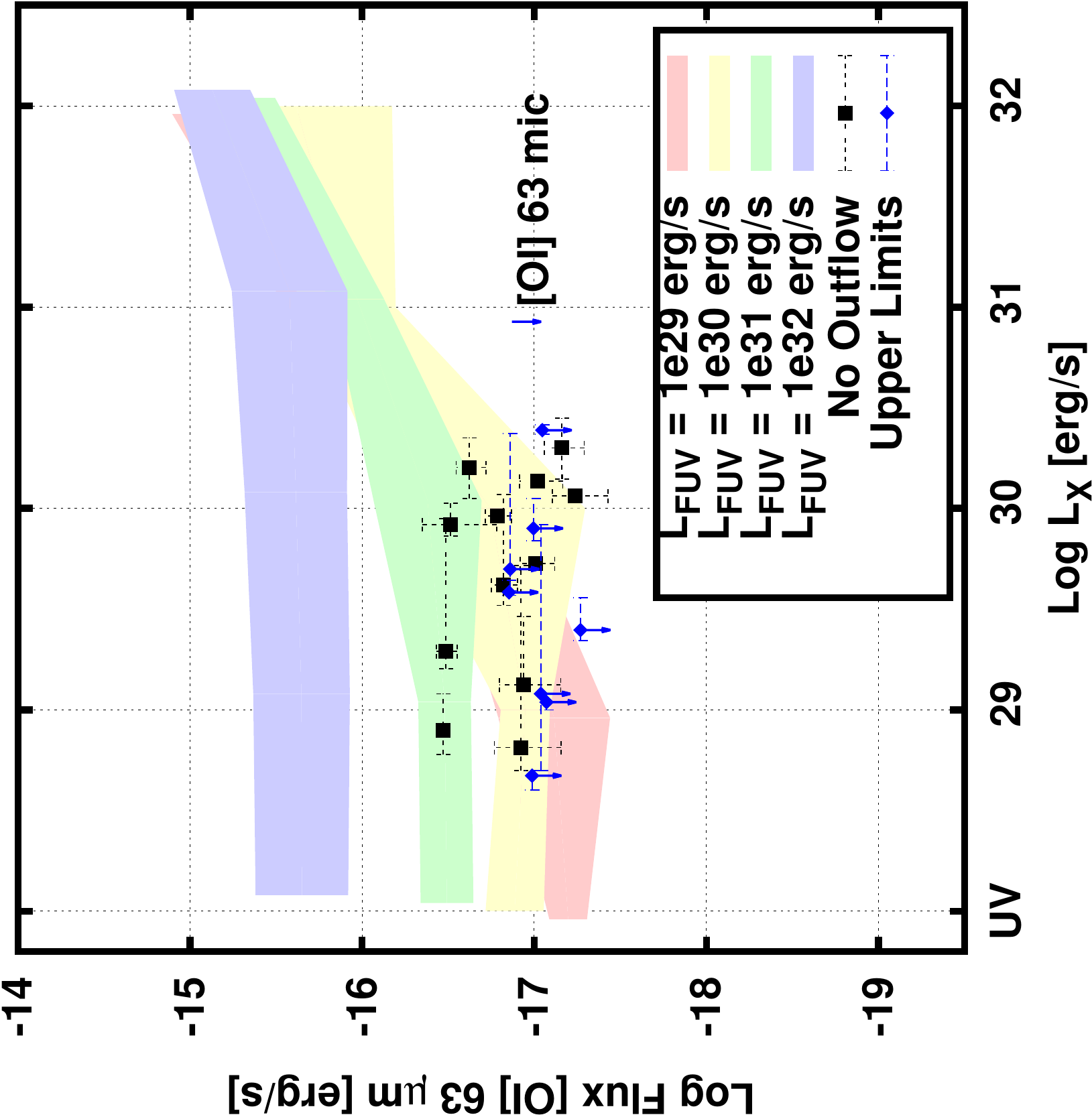}%
 \hspace{0.2cm}
 \includegraphics[scale=0.4,angle=-90]{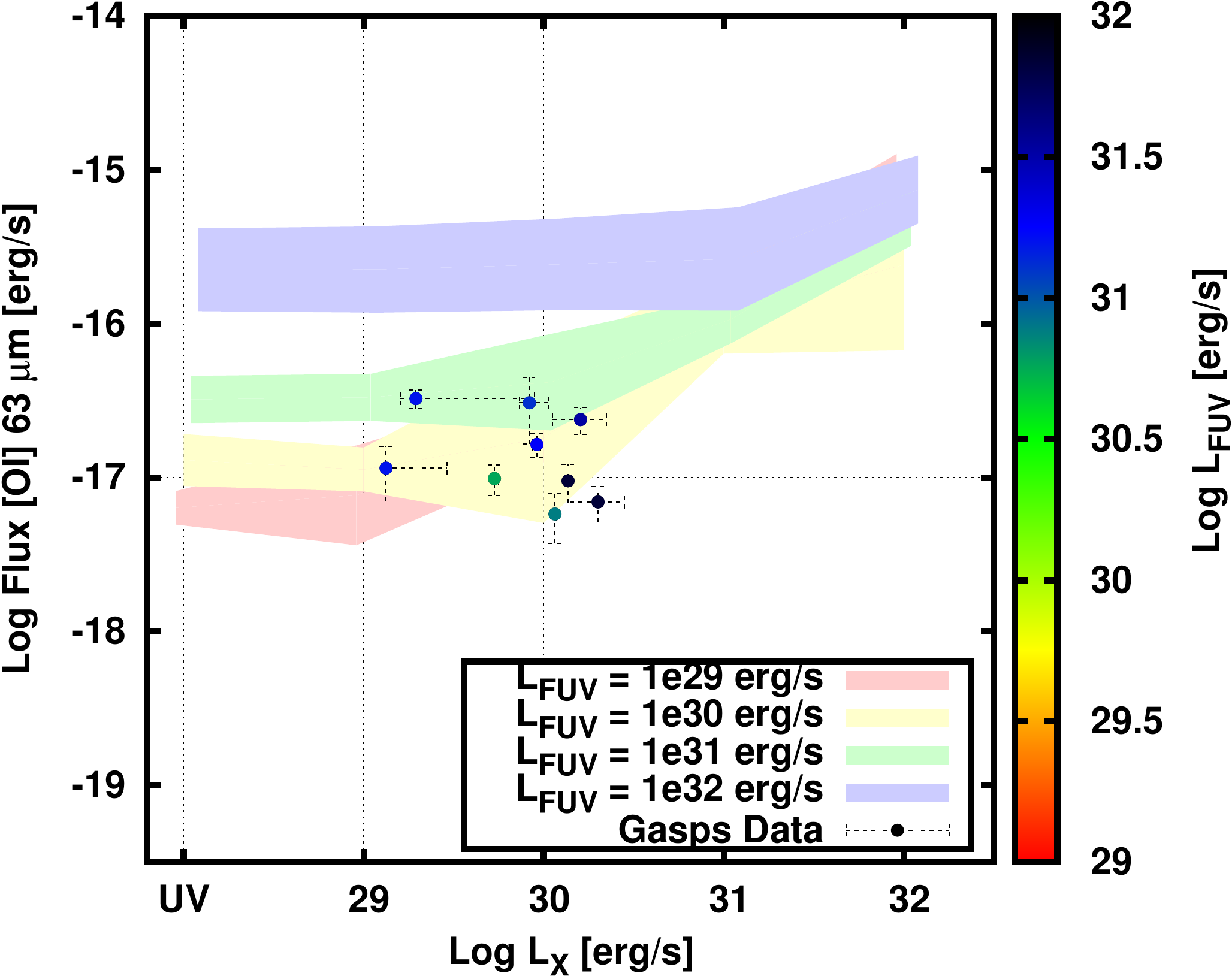}
 \caption{\small Models and observations. FUV luminosity ranges from 10$^{29}$ (red stripe) to 10$^{32}$ erg/s (blue stripe). When stripes overlap, the stripe representing a higher luminosity is shown. In the right-hand panel the data points are color coded for the FUV luminosity of a given object, following the color-scale presented in the side bar.}
 \label{modX}
\end{figure*}

\subsection{Observed data}
In the left-hand panel of Fig. \ref{oilxdata}, we plot the \on\,fluxes versus the X-ray luminosity. Red dots identify disk only sources while green diamonds are upper limits. The X-ray luminosity range spans $\sim 1.5$ dex, as well as the range in \on\,fluxes. Due to the presence of upper limits, we perform survival analysis using the ASURV package \citep{Fei85,Iso86} to investigate quantitatively the presence of a correlation. In Table \ref{asurv}, we summarize the results obtained showing the probability that the correlation is not present using a Spearman and Kendall statistical test. This has been done for the data sample and for a random population of values in the same ranges (in brackets). 

In the right-hand panel of Fig. \ref{oilxdata}, we plot the \on\,fluxes versus \luv. 
Also in this case there is no clear correlation between \on\, and the FUV luminosity. We found similar results concerning the correlation between \on\,and \lsum = \lx +\luv\,(Table \ref{asurv}).

\subsection{Modeling}

\begin{table*}[ht]

\centering
\begin{tabular}{l|c|c}
\hline
\hline
Observable &  Kendall & Spearman \\
\hline 
\lx      & 0.77 (0.60) & 0.80 (0.51) \\
\hline
\luv   & 0.84 (0.41) & 0.81 (0.40) \\
\hline
\lsum   & 0.83   & 0.83 \\
\hline
\end{tabular}
\caption{\small Probability that a correlation is not present estimated using Spearman and Kendall coefficients for the correlation between \on\,and \lx, \luv\,and \lsum (in brackets, the probabilities for a random population in the same x and y range).}
\label{asurv}

\end{table*}

In the left-hand panel of Fig. \ref{modX} we show the results taken from the grid of models described in \citet{Mei12} and \citet{Are12}, each coloured stripe is a series of \on\,fluxes for models with a given FUV luminosity. The thickness of the stripe accounts for all the models with different dust parameters and disk surface density distribution (see \citealt{Are12} for the details). The models agree quantitatively with the data, reproducing the same \on\,flux range from low to high FUV luminosity along the 2 dex interval in \lx. The models do not predict a correlation between \on\,and \lx, but rather a threshold behaviour: for \lx $> 10^{30}$ erg/s and \lx $>$ \luv, \on\,emission should be dominated by X-rays.  

\begin{figure*}[t!]
\centering
 \includegraphics[scale=0.4,angle=-90]{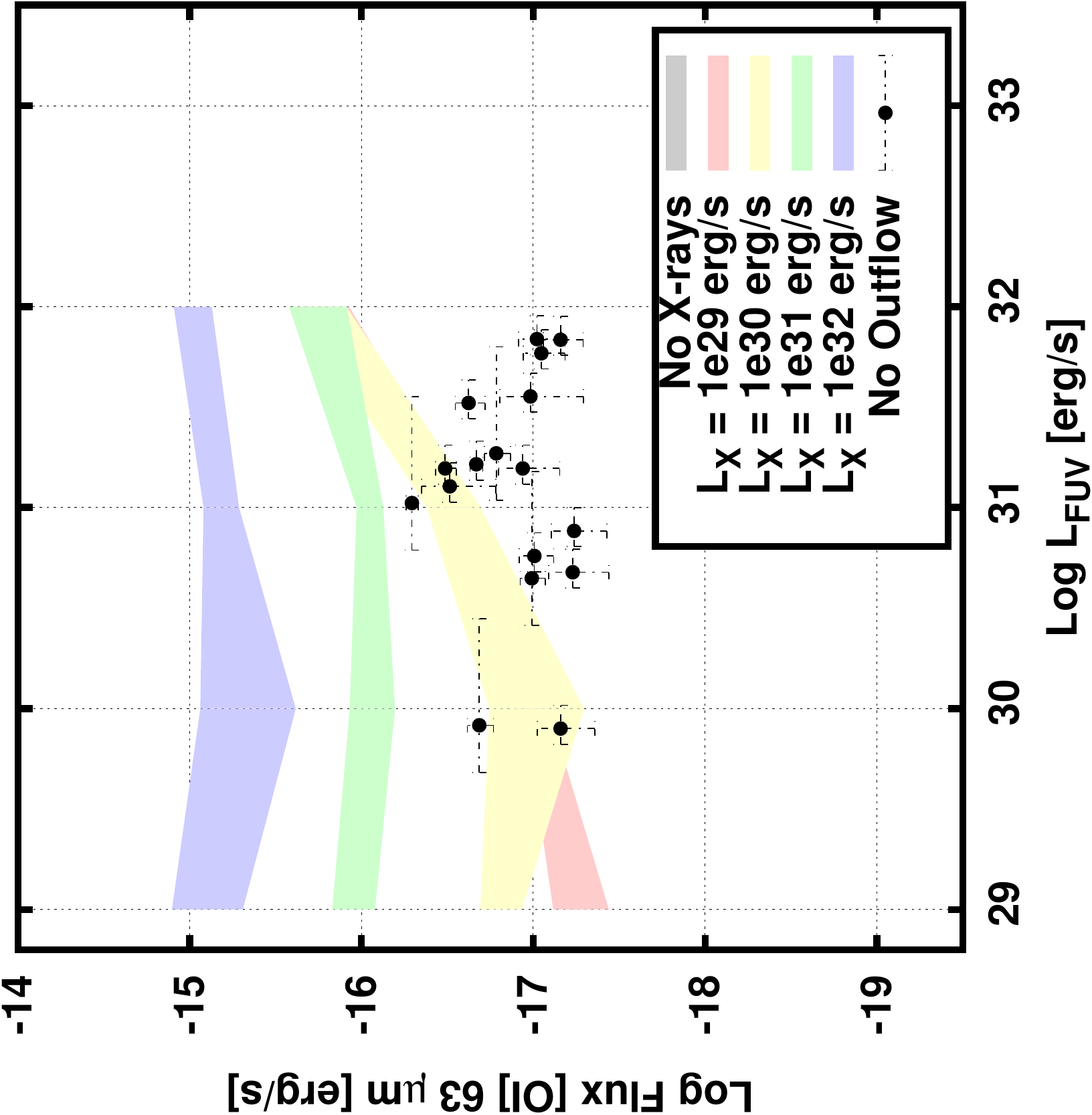}%
 \includegraphics[scale=0.4,angle=-90]{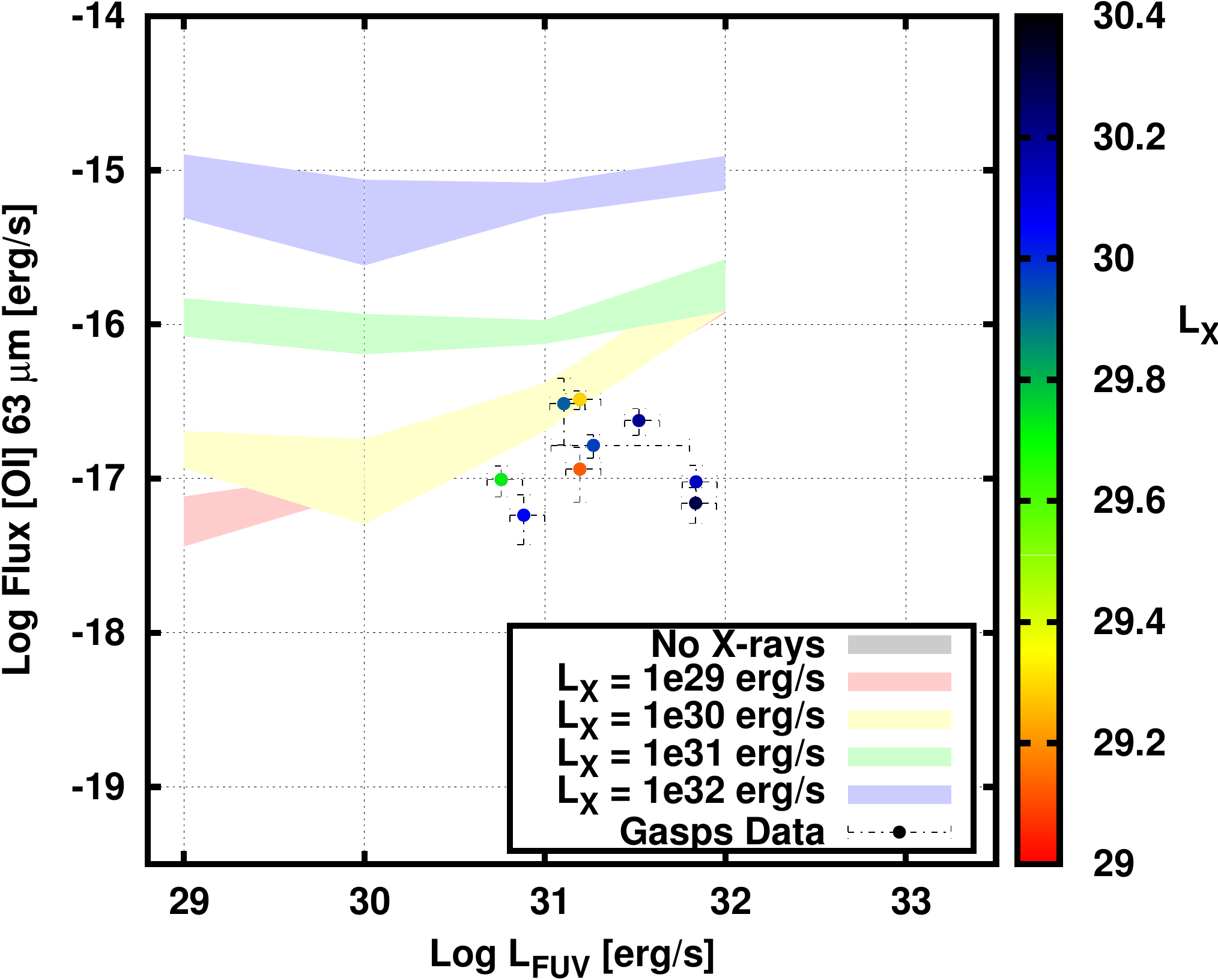}
 \caption{\small Flux of the \on\,63 \mic\,line versus \luv. The coloured stripes represent predicted values for different \lx.}
\label{uvcc}
\end{figure*}

The models also suggest that at a given X-ray luminosity, the \on\,line flux scales with \luv. To test this on a qualitative basis, in the right-hand panel of Fig. \ref{modX}, we plot the \on\,observed line fluxes versus the X-ray luminosity, colour coding for the observed FUV luminosity. The predicted \on\,fluxes from the models seem to overestimate (factor $\sim$5) the observations at a given \luv. This can be also seen in Fig. \ref{uvcc}, where the predicted \on\,is on average higher than the data. In this plot the stripes are colour coded for different \lx. These findings depend strongly on how the FUV luminosity is scaled from the 7-10 eV band to the 6-13.6 band. We use the same stellar template (TW Hya) to estimate the variation of the flux in the full FUV range. However this might not be applicable to each object. Moreover, we do not consider the Ly$\alpha$ flux in our models, which might cause an overestimate of the continuum flux in the FUV band, that could cause extra FUV heating in the \on\,emission region, hence overestimating its line flux.

Fig. 2 in Aresu et al. (2012, right-hand panel), shows that the energy deposition rates associated with \lx\,and \luv\,are comparable. Hence, we explore the existence of a correlation between \on\,and the simple sum of these luminosities. We fit the data using a linear function. Fig. \ref{fit3} shows the data and the fit to the model points (red line). Given that all the sources in the observed sample have \lx\,between 10$^{29}$ and 10$^{31}$ erg/s, we did not include the \on\,flux for those models which have \lx $= 0$ or $ 10^{32}$ erg/s, obtaining a slope of 0.6.

\section{Discussion}
\label{disc}
We discuss here the results presented in the previous section, analysing the capability of the models in interpreting the data and suggest improvements to be made. 

\subsection{FUV luminosity}
\begin{figure}[h!]
\centering
 \includegraphics[scale=0.4,angle=-90]{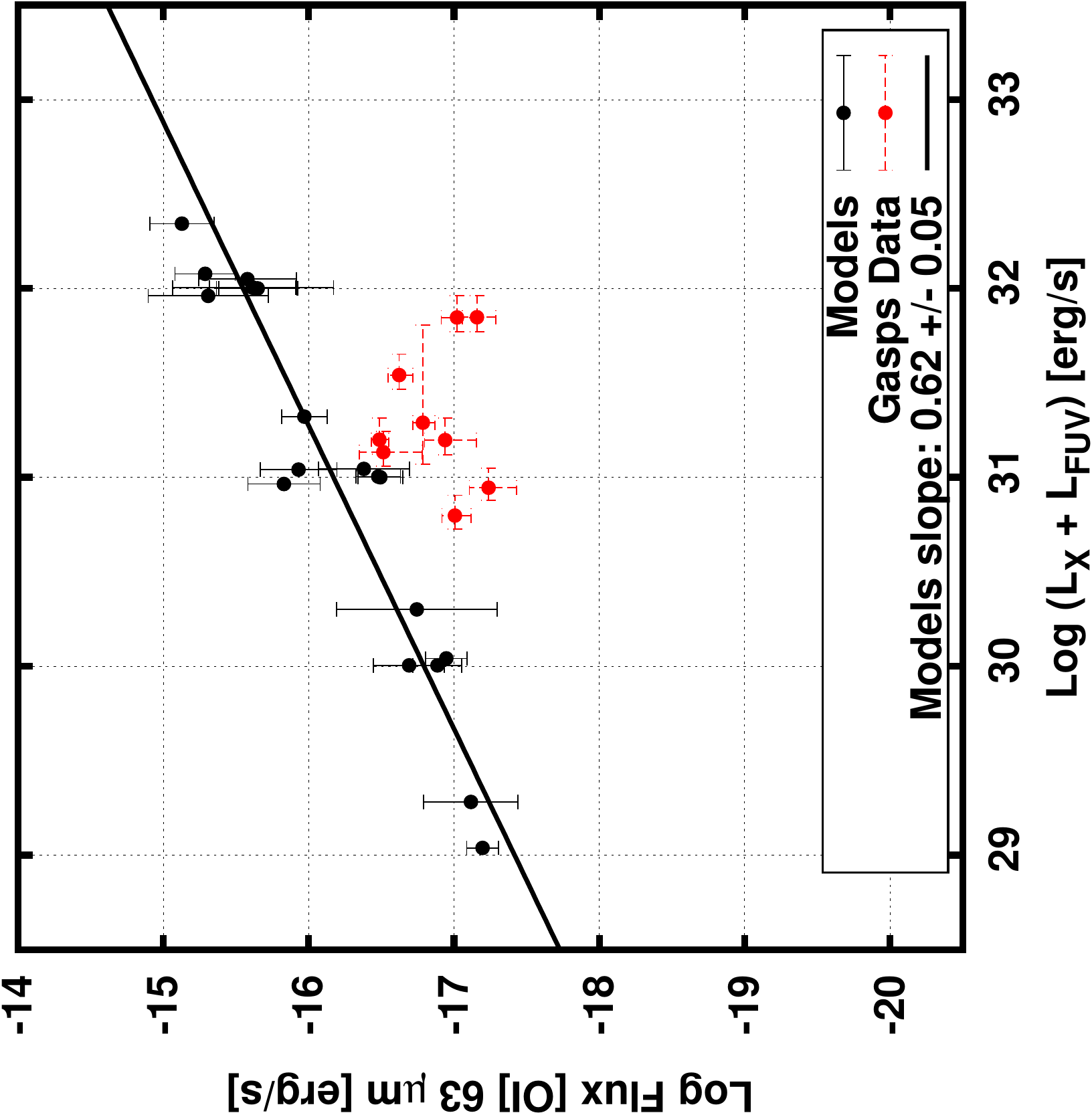}%
 \caption{\small \on\,flux versus \lx+\luv. Model points are black while data points are red. In our models we considered a sun-like star surrounded by a disk of 0.02 $M_{\odot}$ which extends from 0.5 to 500 AU.}
\label{fit3}
\end{figure}

Our \luv\,estimates rely on the calculation of the scale factor that we adopted to get the luminosity in the 6-13.6 band (needed to compare the data to our models) out of the 7-10 eV band used by \citet{Yan12}. These values are subject to improvement for two reasons: (1) we calculated the conversion factor for TW Hya only, and used it for the whole sample, (2) the Ly-$\alpha$ profile in the TW Hya spectrum should be reconstructed to account for neutral hydrogen absorption in the ISM. Nevertheless the slope in Fig. \ref{fit3} is weakly dependent on this reconstruction. A change of a factor 2 in the conversion factor would cause a $\sim$20\% change in the slope.
\subsection{Possible correlations}
Assuming that the measured \on\,fluxes are mainly emitted from the disk, we find that both the models and the data suggest that there is no correlation between \on\,63 \mic\,and \lx\,or \luv\,in the range of luminosities spanned by our sample. According to our models this is caused by the fact that FUV-related heating processes affect the line at a given X-ray luminosity, causing a vertical scatter ($\sim$ 2 dex) comparable with the range of X-ray luminosities observed in T\,Tauri stars. The same conclusions can be drawn when \on\,is plotted against \luv, where the vertical scatter is now to ascribe to Coulomb heating.

This amount of scatter is partially seen in the data ($\sim$1 dex), and no correlation between \on\,and \luv\,emerges. All the sources investigated here have \luv\,$>$\lx, and 90\% of them have \luv $>10\times$\lx. Disk models would predict higher \on\,emission (factor $\sim$10) for \luv$>$10$^{31}$ erg/s, but the data shows that higher \on\,fluxes are achieved only when outflows are present. 

With the aim of extending our data sample, we included also outflow sources and proceeded to estimate the \on\,disk emission as follows: \citet{How13} find a correlation between \on\,and the continuum at 63 $\mu$m for non-outflow sources. Outflow sources stand clearly above this correlation (see their Fig. 6). Using the fit formula that they provided, we estimated the disk emission for outflow sources subtracting from the total \on\,emission the amount of excess flux with respect to the fit (table \ref{sample}, fourth column). In this way we enlarge our sample (though only through estimated disk emission) and attempt to check the absence of a clear correlation with \luv\,and \lx. In fact, even for the enhanced sample, we find no correlation with either one of these quantities, nor with their sum.
\subsection{Heating mechanism}
Since the PACS data is spatially and spectrally unresolved, the location of the \on\,63 \mic\,emission is unclear. If most of the emission for non-outflow sources originates in the disk, both \lx\,and \luv\,are important heating agents in the \on\,emitting region, but their direct influence on the \on\,emission remains elusive. In \citet{Are12}, we suggest a threshold mechanism for \on\,with respect to \lx. This cannot be tested with our current data-set, as there are no sources that have \lx\,higher than few times 10$^{30}$ erg/s.

\subsection{Impact of other parameters}
Spitzer observations toward 38 T\,Tauri stars performed by \citet{Ger06}, detected PAH features in only 8\% of the objects. Models of PAH chemistry in disks surrounding T\,Tauri stars suggest that these species do exist, but the UV luminosity of the central star is just too weak to reveal their presence. PAH emission is indeed believed to be a factor 10 weaker in T\,Tauri stars when compared to Herbig Ae/Be stars, where the UV luminosity is orders of magnitude higher. The authors also suggest that the PAH abundance in T\,Tauri stars is a factor 10 or 100 lower than the one inferred for the ISM.

In our grid we used a PAH abundance 0.012 times lower than the PAH abundance in the ISM. Nevertheless, we found PAH heating to be the main FUV driven heating process in the \on\,emission region. The second most efficient heating processes are C$^+$ heating and photoelectric heating on dust grains. The sum of these heating processes is less than a factor two lower than PAH heating, and follows the same behaviour with respect to \luv. An even lower PAH abundance would just cause a lower temperature in the \on\,emission region, thus weakening the line emission, but not affecting the nature of the correlation we predict.

On the other hand we considered in our models only a limited set of free parameters (\lx, \luv, minimum dust size, dust size distribution power law and the surface density distribution power law). However the \on\,flux can be affected by other quantities, e.g. the flaring index $\beta$, dust settling, gas-to-dust mass ratio and outer radius \citep{Woi10,Kam11}. In the disk models used in this work, the flaring angle  is a result of the hydrostatic equilibrium \citep{Mei12} and the solutions we find for the scale height generally point toward maximum flared disks ($\beta\sim$1.25). This may not be representative of the disks in our sample especially when dust settling takes place, leading to flatter geometries, for which $\beta \leq$1 \citep{Dul05}. Flat disks absorb less radiation hereby diminishing the importance its importance, hence its impact on the gas emission.

Dust settling should also affect the dust-to-gas mass ratio which is usually kept fixed to the ISM value ($\delta$=0.01) at each point in the disk. Variations of the latter would change the opacity throughout the disk causing different properties in the energy deposition distribution of the FUV radiation.

In our models we keep the disk outer radius fixed at 500 AU to allow proper comparison with previous works. However disks surrounding T\,Tauri stars do range in size from 50 AU to hundreds of AU \citep{Wil11}, e.g. due the reduced emitting area, smaller disks illuminated by high FUV luminosities not necessary yield higher \on\,fluxes when compared to bigger disks illuminated by lower \luv.

The disk mass also affects the \on\,emission: in Taurus the spread in disk masses is estimated to be $\sim$2 dex, being M$_D/$M$_{\odot} \sim10^{-1.5}-10^{-3.5}$ \citep{And13}. This spread could cause $\sim$ 1 dex scatter in the \on\,emission \citep{Woi11}, hence affecting the correlations studied in this work.

\section{Conclusions and outlook}
\label{fut}
In this work we studied the impact of FUV and X-ray radiation on the thermal balance in the oxygen emission region for protoplanetary disks surrounding T\,Tauri stars. We compared disk model predictions with observations of the \on\,63\mic\,line toward protoplanetary disks that do not show outflow emission in the Taurus region obtained with the PACS instrument on board Herschel. 

The observations show no correlation between the \on\,63 \mic\,line emission and the X-ray luminosity or the FUV luminosity or with their sum.

Our thermo-chemical disk models calculated with ProDiMo, show that our predictions on the \on\,fluxes qualitatively agree with the observations. There is no correlation between \on\,and \lx\,or \luv, as the data suggest. Nevertheless, the models predict a correlation between \on\,and the sum \lx+\luv, which is not seen in the data. The reason can be the limited set of parameters varied in our model (\lx, \luv, grain minimum size, power law of the grain size distribution and power law of the surface density distribution) grid to understand the relative importance of \lx\,and \luv.However, other parameters can affect the \on\,line, causing the correlation we predict to vanish when a more complete grid is used.

To include all the disk parameters that influence the \on\,line, a different set of models should be used. Flatter disk geometries should be included as well as a proper treatment of X-ray and FUV physics and dust settling (local variations of the gas-to-dust mass ratio).

Moreover, such models should be compared to a higher number of observations: high spatial and spectral resolution data is required to disentangle the location of the emission region of the line. In many sources that drive outflows, the contribution of the disk to the total flux of the line remains unclear.

More measurements of \luv\,would also be necessary. To test the threshold mechanisms proposed in our previous work, observations of \on\,of sources with \lx $>$ \luv, if any, are essential. 

The understanding of the [OI] dependence on the FUV and X-ray radiation gives the possibility to investigate the gas surface layers above the H/H2 transition. Such studies are very interesting for understanding the photoevaporation mechanism and how it may drive disk evolution across the transition from optically thick to debris disk. 

\tiny \emph{Acknowledgements}.  We thank Aki Roberge for her comments on the FUV analysis which helped to improved the paper and Glenn White for thoroughly reading the paper. WFT, PW, FM, MG and IK acknowledge funding from the EU FP7-2011 under Grant Agreement nr. 284405. L.P. acknowledges the funding from the FP7 Intra-European Marie Curie Fellowship (PIEF-GA-2009-253896).

\bibliographystyle{aa}
\bibliography{biblio}
\end{document}